\documentstyle[seceq,preprint,mbf,epsf]{ptptex}

\title{
Asymmetry in Microlensing-Induced Light Curves
}

\author{
Kohkichi {\sc Konno} 
and Yasufumi {\sc Kojima}
}

\inst{
Department of Physics, Hiroshima University, 
Higashi-Hiroshima 739-8526, Japan
}

\recdate{
December 8, 1998%
}

\abst{

We discuss distortion in microlensing-induced light curves
which are considered to be curves due to single-point-mass 
lenses at a first glance. 
As factors of the distortion, we consider 
close binary and  planetary systems, which are
the limiting cases of two-point-mass lenses, and 
the gravitational potential is regarded as the sum of 
the single point mass and the corrections. In order to  
quantitatively estimate
the asymmetric features of such distorted light
curves, we propose a cutoff dependent 
skewness, and show how to  discriminate 
the similar light curves with it.
We also examine as the  distortion
the general relativistic effect of frame dragging, 
but the effect coincides with the close binary case in the 
light curves.
}

\notypesetlogo

\begin{document}

\maketitle

\section{Introduction}

Gravitational microlensing \cite{microl} is one important
probe for 
studying the nature and distribution of mass in the galaxy.
Chang and Refsdal \cite{cr} and Gott \cite{gott} suggested
that even though  multiple images by lensing are 
unresolved,  the time variation of the magnitude of 
the source can still be detected if the lens moves relative
to the source. Such light curves caused by the microlensing
can be distinguished from curves of intrinsically variable 
sources, because the change of magnitude by lensing is
achromatic, whereas the colors of intrinsically variable stars change 
in general. Furthermore, microlensing-induced
light curves can be distinguished 
from magnification by bursts, which are likely to appear
in sheer shapes. 
As is well known, microlensing by a point mass
has the time-symmetric light curves, provided that the lens
and the source have constant relative transverse velocity.
Hence, most events are expected to have almost time-symmetric
light curves. If the relative velocity is not constant,
then of course time-asymmetric light curves will
be detected. Gould \cite{gould} predicted a parallax effect
due to the orbital motion of the Earth, and its effect was indeed
detected by Alcock el al.\cite{alcock} \ 
When the time scale of a microlensing event
is larger than $\sim 100$ days, this effect is important.
However, if the time scale is of hours
to weeks, the parallax effect can be neglected.
Then, other factors, 
such as the non-spherical gravitational potential of the lens,
could also produce distortion from 
the time-symmetric light curves.
The effect may be regarded as a 
higher-order correction to the point-mass lens. 
The aim of this paper is 
to evaluate the effect of the intrinsic nature of the lenses,
which may slightly distort the light curves. 
We exclude significantly peculiar light curves,
such as double peaks, from which direct 
information is available without any detailed analyses.
We rather restrict our consideration to light curves 
which are regarded as curves of single-point-mass lenses at 
a first glance, that is,  almost time-symmetric light curves.
Since the asymmetric part contains additional
information about massive astrophysical compact halo
objects (MACHOs), this subject is very important
to understanding the nature of MACHOs.

One of the important factors to induce 
time-asymmetric forms is
the binary system of the lensing objects, in which 
the contribution from both objects to 
the Newtonian gravitational potential is no longer spherical.
The discussion is simplified by
considering two-point-mass lenses.
In particular, since we consider almost time-symmetric 
light curves, we hit on two situations: 
close binary and planetary systems.
In a close binary system, the separation
distance between the two point masses is much less than
the Einstein radius of the total mass, so that
the approximated light curve can be described by
the total mass and some corrections to it.
The detectability of a close binary system has been
discussed in detail by Gaudi and Gould\cite{gg}
considering the excess magnification threshold.
According to them, the detectability of close
binaries with separation less than $\sim 0.2$
of the Einstein radius is $\sim 10\%$.
Therefore, most close binary lenses with 
such small separation are missed. We consider
the possibility of picking up the discarded asymmetry, by which
even below the separation of $\sim 0.2$ of the
Einstein radius the binary nature of the lenses may be detected.
In a planetary system, the 
light curve is expected to be described by the contribution 
from the larger mass and some corrections from the smaller one.
Several authors\cite{mp,gl,bf,br,gg} have discussed planetary
systems with remarkable deviations from a single lens.
However, we are now interested in planetary events which
would be missed due to the special configurations of the 
lensing geometry. Our new proposal may make the detection
of such events possible, in addition to the close binary case.
Another factor for the time-asymmetric forms is
the asymmetry of the lens object itself.
The multipole moment of the lens deviates from gravitational
potential of the point mass. The  gravitational
potential due to the quadrupole never vanishes  for
the two-point-mass case. Therefore, such a 
correction term in the gravitational  potential
can be regarded as the limiting case of the binary.
Instead, we shall consider  general relativistic
effects of dragging of inertial frames
due to a rotating object as another factor,
which cannot be expressed as corrections to the
Newton-like potential. 
While the Newton-like potential corresponds to the
gravitoelectric field, this effect results from
the gravitomagnetic field.
In this case, corrections up 
to post-Newtonian order are sufficient.
Other post-Newtonian potentials also affect the 
light curves, but they are neglected here.

This paper is organized as follows. In \S \ref{two-l}
we discuss the two situations involving the two-point-mass
lenses and obtain microlensing-induced 
light curves slightly deviating from the time-symmetric
curves. In \S \ref{kerr} we discuss the post-Newtonian
corrections due to the rotation and obtain
time-asymmetric light curves as well. 
We propose a certain tool to estimate the time-asymmetric
features quantitatively in \S \ref{n-estimates}.
Finally, we give summary and discussion
in \S \ref{s-d}.

\section{Two-point-mass lenses}
\label{two-l}

We investigate two extreme situations involving two-point-mass 
lenses in order. (The general computational details including
critical lines and caustics can be found in  
Ref.~\citen{sw}.) First, we consider the situation 
in which the distance $l$ between the two point masses is much less 
than the Einstein radius $r_{E}$ corresponding to the total mass.
Next, we consider the situation in which one mass, $M_{1}$,
is much smaller than the other mass, $M_{2}$.

\subsection{Close binary system}
\label{l-r}

\begin{figure}[b]
    \epsfxsize= 8 cm
    \centerline{\epsfbox{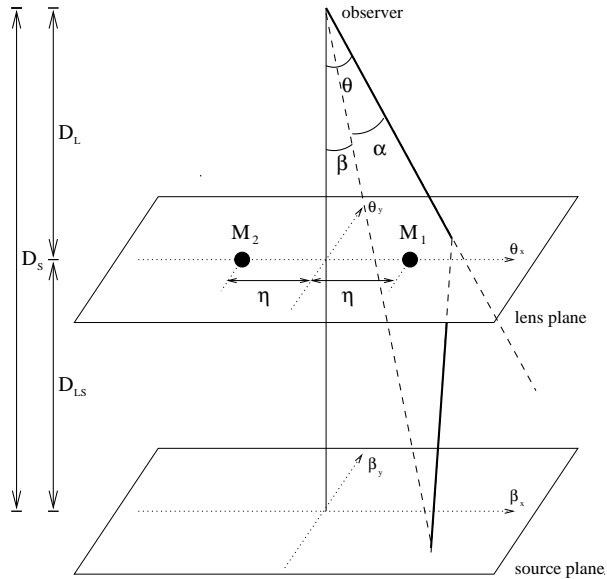}}
    \caption{Geometry of the gravitational lensing considered in \S 2.1.}
    \label{fig1}
\end{figure}

We consider the lens plane on to which the positions of  
the two point masses $M_{1}$ and $M_{2}$ are projected 
(see Fig.~\ref{fig1}).
We introduce an angular coordinate system 
$\left( \theta_{x}, \theta_{y} \right)$,
in which the two point masses lie on the 
$\theta_{x}$-axis and 
the origin is chosen as their geometrical center.
We also define 
an angular coordinate system $\left( \beta_{x}, \beta_{y} \right)$
in the source plane, corresponding to the lens plane.  
We express the distance between
the observer and the lens plane, the lens plane and the source
plane, and the observer and the source plane by $D_{L}$, $D_{LS}$
and $D_{S}$, respectively.

If we denote the angular separation between the source
and the image by ${\mbf \alpha}$, we have
\begin{subequations}
\begin{eqnarray}
 \alpha_{x}
 &=&\frac{4GM_{1}}{c^2} \frac{D_{LS}}{D_{L} D_{S}}
       \frac{\theta_{x} - \eta}{\left( \theta_{x}
       - \eta \right)^2 + \theta_{y}^2}
       +  \frac{4GM_{2}}{c^2} \frac{D_{LS}}{D_{L} D_{S}}
       \frac{\theta_{x} + \eta}{\left( \theta_{x}
       + \eta \right)^2 + \theta_{y}^2} , \quad \\
 \alpha_{y}
 &=& \frac{4GM_{1}}{c^2} \frac{D_{LS}}{D_{L} D_{S}}
       \frac{\theta_{y}}{\left( \theta_{x}
       - \eta \right)^2 + \theta_{y}^2}
       +  \frac{4GM_{2}}{c^2} \frac{D_{LS}}{D_{L} D_{S}}
       \frac{\theta_{y}}{\left( \theta_{x}
       + \eta \right)^2 + \theta_{y}^2} ,
\end{eqnarray}
\end{subequations}
where $\eta$ is the angular separation of the mass
$M_{1}$ (or $M_{2}$) from the optical axis. 
Therefore, the lens equation can be written in the form
\begin{equation}
 {\mbf \beta} = {\mbf \theta} - {\mbf \alpha} .
\end{equation}
We normalize this equation by the Einstein radius for 
the total mass,
\begin{equation}
 \theta_{E} = \left( \frac{4G \left( M_{1} + M_{2} \right)}
   {c^2} \: \frac{D_{LS}}{D_{L} D_{S}}\right)^{\frac{1}{2}} ,
\end{equation}
and introduce normalized quantities
\begin{equation}
 \tilde{\mbf \beta} \equiv \frac{{\mbf \beta}}{\theta_{E}},
   \qquad \tilde{\mbf \theta} \equiv \frac{{\mbf \theta}}
   {\theta_{E}}, \qquad \tilde{\eta} \equiv 
   \frac{\eta}{\theta_{E}} .
\end{equation}
The lens equation is then given by
\begin{subequations}
\begin{eqnarray}
  \tilde{\beta}_{x} 
  & = & \tilde{\theta}_{x} - \mu_{1} \frac{\tilde{\theta}_{x} 
        - \tilde{\eta}}{\left( \tilde{\theta}_{x}
        - \tilde{\eta} \right)^2 + \tilde{\theta}_{y}^2}
        - \mu_{2} \frac{\tilde{\theta}_{x} 
        + \tilde{\eta}}{\left( \tilde{\theta}_{x}
        + \tilde{\eta} \right)^2 + \tilde{\theta}_{y}^2} ,\\
  \tilde{\beta}_{y} 
  & = & \tilde{\theta}_{y} - \mu_{1} \frac{\tilde{\theta}_{y}} 
       {\left( \tilde{\theta}_{x}
        - \tilde{\eta} \right)^2 + \tilde{\theta}_{y}^2}
        - \mu_{2} \frac{\tilde{\theta}_{y}} 
        {\left( \tilde{\theta}_{x}
        + \tilde{\eta} \right)^2 + \tilde{\theta}_{y}^2} ,
\end{eqnarray}
\end{subequations}
where $\mu_{1}$ and $\mu_{2}$ are defined as
\begin{equation}
  \mu_{1} = \frac{M_{1}}{M_{1}+M_{2}} , \qquad
  \mu_{2} = \frac{M_{2}}{M_{1}+M_{2}} .
\end{equation}

The separation  distance in the projected plane is
$l = D_{L} \cdot 2 \eta$, and
the Einstein radius for  this system
$r_{E} = D_{L} \cdot \theta_{E}$. \ 
The term `close binary' in the gravitational lens 
means $l  \ll r_{E}.$ \ 
This condition in the astronomical situation
is expressed as
$ l  \ll 10^{14}(M/M_\odot)^{1/2}$ ($D$/(10 kpc))$^{1/2}$ cm,
where we have chosen typical astronomical distances as
the scale of the galactic halo,
$ D_{LS} \sim D_{L} \sim D_{S} \sim D .$ \ 
Therefore, the range of applicability is not so severely limited.
The condition of the close binary,  
$l  \ll r_{E}$, is mathematically expressed as 
\begin{equation}
  \tilde{\eta} \ll 1.
\end{equation} 

Under this condition, we expand the right-hand side of the 
lens equation with respect to $\tilde{\eta}$. Up
to first order in $\tilde{\eta}$, we have
\begin{subequations}
\label{lens-eq-1a}
\begin{eqnarray}
  \tilde{\beta}_{x} 
  & = & \tilde{\theta}_{x} - \frac{\tilde{\theta}_{x}}
        {\tilde{\theta}_{x}^{2} + \tilde{\theta}_{y}^2}
        - \tilde{\eta} \left( \mu_{1} - \mu_{2} \right)
        \frac{\tilde{\theta}_{x}^{2} - \tilde{\theta}_{y}^{2}}
        {\left( \tilde{\theta}_{x}^2
        + \tilde{\theta}_{y}^2 \right)^2}  ,\\
  \tilde{\beta}_{y} 
  & = & \tilde{\theta}_{y} - \frac{\tilde{\theta}_{y}} 
       {\tilde{\theta}_{x}^2 + \tilde{\theta}_{y}^2}
        - \tilde{\eta} \left( \mu_{1} - \mu_{2} \right) 
        \frac{2 \tilde{\theta}_{x} \tilde{\theta}_{y}} 
        {\left( \tilde{\theta}_{x}^2 + \tilde{\theta}_{y}^2
        \right)^2} ,
\end{eqnarray}
\end{subequations}
where the last term on each right-hand side represents the deviation
from a single-point-mass lens. 
The first-order corrections vanish for the equal mass case,
since $\mu_{1} = \mu_{2}$.
It is therefore convenient to express the 
combination $\tilde{\eta} \left( \mu_{1} - \mu_{2} \right)$
as one small parameter 
$\varepsilon \equiv \tilde{\eta} \left( \mu_{1} - \mu_{2} \right)$.

The inversion of Eq.~(\ref{lens-eq-1a}), i.e., 
solving $\tilde{\theta}$ by $ \tilde{\beta} $, 
is possible, but the general form is quite messy.
However, under the condition $\varepsilon \ll 1$, 
the approximate solution, i.e., 
the first-order solution in $\varepsilon$,
is given by the form
\begin{equation}
\label{find-r}
 \tilde{\mbf \theta} 
   = \tilde{\mbf \theta}_{0} 
      + \varepsilon \: \tilde{\mbf \theta}_{1} ,
\end{equation}
where $\tilde{\mbf \theta}_{0}$ and
$\tilde{\mbf \theta}_{1}$ denote the zeroth-order and the 
first-order solutions, respectively. Substituting
Eq.~(\ref{find-r}) into Eqs.~(\ref{lens-eq-1a}), 
we can find such solutions. The
zeroth-order solution is given by
\begin{subequations}
\label{zero-s1}
\begin{eqnarray}
 \tilde{\theta}_{0x} 
 & = & \frac{1}{2} \left( \tilde{\beta}_{x} \pm
       \tilde{\beta}_{x} \sqrt{1 + \frac{4}
       {\tilde{\beta}_{x}^{2} + \tilde{\beta}_{y}^{2}}} \right) , \\
 \tilde{\theta}_{0y} 
 & = & \frac{1}{2} \left( \tilde{\beta}_{y} \pm
       \tilde{\beta}_{y} \sqrt{1 + \frac{4}
       {\tilde{\beta}_{x}^{2} + \tilde{\beta}_{y}^{2}}} \right) .
\end{eqnarray}
\end{subequations}
Using this zeroth-order solution, the first-order
solution is written in the form
\begin{subequations}
\begin{eqnarray}
 \tilde{\theta}_{1x}
 & = & \frac{\tilde{\theta}_{0x}^{\: 2} - \tilde{\theta}_{0y}^{\: 2}-1}
       {\left( \tilde{\theta}_{0x}^{\: 2} + \tilde{\theta}_{0y}^{\: 2} 
        \right)^{2} - 1} , \\
 \tilde{\theta}_{1y}
 & = & \frac{2 \tilde{\theta}_{0x} \tilde{\theta}_{0y}}
       {\left( \tilde{\theta}_{0x}^{\: 2} + \tilde{\theta}_{0y}^{\: 2} 
        \right)^{2} - 1} .
\end{eqnarray}
\end{subequations}

\begin{figure}[t]
    \epsfxsize= 8 cm
    \centerline{\epsfbox{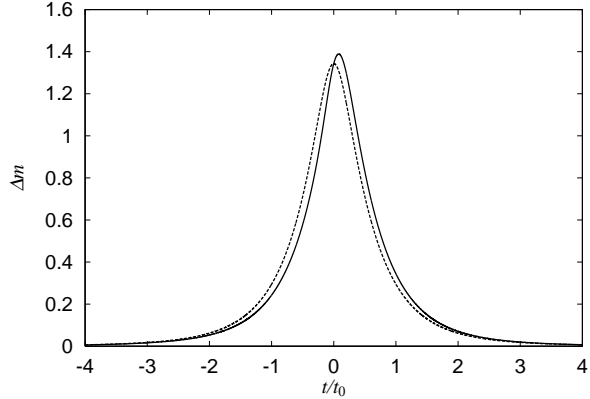}}
    \caption{Microlensing-induced light curves with (solid) and
      without (dashed) correction for the deviation from a 
      single-point-mass lens. The light curve with the correction
      is plotted for the small parameter value $\varepsilon = 0.1$.
      The relative motion of the source to the lens is assumed
      to be described by $\varphi = 0$ and $p=0.3$.}
    \label{graph1}
\end{figure}

Next, we turn our attention to the derivation of magnification.
The magnification $M$ is given by
\begin{eqnarray}
\label{mag}
 M & = & M_{+} + M_{-} \nonumber \\ 
 & = & \left| \det \left( \frac{\partial 
       \tilde{\beta}_{i}}{\partial \tilde{\theta}_{j}} \right)_{+}
       \right|^{-1} + \left| \det \left( \frac{\partial
       \tilde{\beta}_{i}}{\partial \tilde{\theta}_{j}} \right)_{-}
       \right|^{-1} ,
\end{eqnarray}
where the subscript $(+)$ and $(-)$ correspond to solutions
with a plus sign and with a minus sign, respectively, in 
Eqs.~(\ref{zero-s1}). The inverse of the Jacobian 
$\det \left( \frac{\partial \tilde{\beta}_{i}}
{\partial \tilde{\theta}_{j}}\right)$ is calculated, up
to first order in $\varepsilon$, in the following way:
\begin{eqnarray}
\label{jacobian}
 \left[ \det \left( \frac{\partial \tilde{\beta}_{i}}
   {\partial \tilde{\theta}_{j}}\right)\right]^{-1}
 & = & \left[ 1 - \frac{1}{\left( \tilde{\theta}_{0x}^{2}
       + \tilde{\theta}_{0y}^{2} \right)^{2}}
       - \varepsilon \frac{4 \tilde{\theta}_{0x}}
       {\left( \tilde{\theta}_{0x}^{2} 
       + \tilde{\theta}_{0y}^{2} \right)^{3}}
       + \varepsilon \frac{4 \left( \tilde{\theta}_{0x} \tilde{\theta}_{1x}
       + \tilde{\theta}_{0y} \tilde{\theta}_{1y} \right)}
       {\left( \tilde{\theta}_{0x}^{2} 
       + \tilde{\theta}_{0y}^{2} \right)^{3}} \right]^{-1}
       \nonumber \\
 & = & \left( 1 - \frac{1}{\left( \tilde{\theta}_{0x}^{2}
       + \tilde{\theta}_{0y}^{2} \right)^{2}} \right)^{-1}
       \left[ 1 + \varepsilon \frac{4 \tilde{\theta}_{0x}}
       {\left( \tilde{\theta}_{0x}^{2} 
       + \tilde{\theta}_{0y}^{2} + 1 \right)^{2}
       \left( \tilde{\theta}_{0x}^{2} 
       + \tilde{\theta}_{0y}^{2} - 1 \right)} \right] .
       \nonumber \\
\end{eqnarray}
Therefore, using Eqs.~(\ref{zero-s1}),
(\ref{mag}) and (\ref{jacobian}), we can derive the 
magnification as a function of the source position:
\begin{equation}
 M = M \left( \beta_{x}, \beta_{y}; \varepsilon \right) .
\end{equation}
The time variation is produced by the relative change of 
the positions. 
The time variation of the magnitude $\Delta m $ is then given by
\begin{equation}
 \Delta m = \Delta m (t) = 2.5 \log_{10} M 
    \left( \beta_{x}(t) , \beta_{y}(t) ; \varepsilon \right) ,
\end{equation}
where the relative source trajectory is in general described
by
\begin{subequations}
\label{s-traj}
\begin{eqnarray}
 \tilde{\beta}_{x}(t) & = & \frac{t}{t_{0}} \cos \varphi 
                        + p \sin \varphi , \\
 \tilde{\beta}_{y}(t) & = & \frac{t}{t_{0}} \sin \varphi 
                        - p \cos \varphi ,
\end{eqnarray}
\end{subequations}
where $t_{0}$ is the time taken to cross the Einstein radius,
$\varphi$ is the angle of the trajectory from the $\beta_{x}$-axis,
and $p$ is the impact parameter normalized by the Einstein radius.
From this, we can derive the microlensing-induced light
curves, slightly deviating from the time-symmetric curves.
An example of the light curves is shown in Fig.~\ref{graph1}
for $\varphi=0$ and $p=0.3$.
The solid line denotes the 
light curve with the correction for the deviation from a
single-point-mass lens, while the dashed line corresponds to
the single-point-mass lens. 
The more general dependence on the angle $\varphi$ and the impact
parameter $p$ is discussed in detail in \S \ref{n-estimates}.

\subsection{Planetary system}
\label{ps}

\begin{figure}[t]
    \epsfxsize= 6 cm
    \centerline{\epsfbox{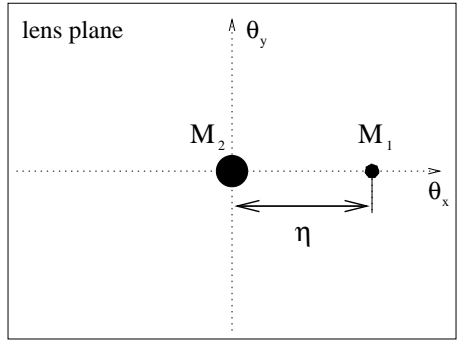}}
    \caption{The lens plane considered in \S 2.2.
        The angular separation of the two point masses
        is denoted by $\eta$.}
    \label{fig2}
\end{figure}

In this subsection, we discuss the case that the mass $M_{1}$
is much smaller than $M_{2}$; that is,
the object with smaller mass is regarded as the planet.
In this situation,
it is convenient to chose the origin of the 
$\left( \theta_{x}, \theta_{y} \right)$ system to be at the position of
mass $M_{2}$. Furthermore, let the angular separation between the
mass $M_{1}$ and $M_{2}$ be $\eta$ (see Fig.~\ref{fig2}).
Then the normalized lens equation is given by
\begin{subequations}
\begin{eqnarray}
 \label{lens-eq-2a}
  \tilde{\beta}_{x} 
  & = & \tilde{\theta}_{x} - \frac{\tilde{\theta}_{x}}
        {\tilde{\theta}_{x}^{2} + \tilde{\theta}_{y}^2}
        - \mu \frac{\tilde{\theta}_{x} - \tilde{\eta}}
        {\left( \tilde{\theta}_{x} - \tilde{\eta} \right)^2
        + \tilde{\theta}_{y}^2}  ,\\
\label{lens-eq-2b}
  \tilde{\beta}_{y} 
  & = & \tilde{\theta}_{y} - \frac{\tilde{\theta}_{y}} 
       {\tilde{\theta}_{x}^2 + \tilde{\theta}_{y}^2}
        - \mu \frac{\tilde{\theta}_{y}} 
        {\left( \tilde{\theta}_{x}^2 - \tilde{\eta} \right)^{2}
         + \tilde{\theta}_{y}^2} ,
\end{eqnarray}
\end{subequations}
where $\mu = M_{1}/M_{2} \left( \ll 1 \right)$, and we have
used for the normalization the Einstein radius of the mass $M_2$,
\begin{equation}
 \theta_{E} = \left( \frac{4G M_{2}}
   {c^2} \: \frac{D_{LS}}{D_{L} D_{S}}\right)^{\frac{1}{2}} . 
\end{equation}
As in the previous subsection,
in the situation $\mu \ll 1$, we approximate
the solutions of the 
lens equation up to first order in $\mu$. Using
the zeroth-order solution (\ref{zero-s1}),
we obtain the first-order solution
\begin{subequations}
\begin{eqnarray}
 \tilde{\theta}_{1x}
 & = & \frac{\left[ \left( \tilde{\theta}_{0x}^{2} 
       + \tilde{\theta}_{0y}^{2} \right)^{2}
       - \left( \tilde{\theta}_{0x}^{2} - 
       \tilde{\theta}_{0y}^{2} \right) \right]
       \left( \tilde{\theta}_{0x} - \tilde{\eta} \right)
       - 2 \tilde{\theta}_{0x} \tilde{\theta}_{0y}^{2}}
       {\left[ \left( \tilde{\theta}_{0x}^{2} 
       + \tilde{\theta}_{0y}^{2} \right)^{2} - 1 \right]
       \left[ \left( \tilde{\theta}_{0x} - \tilde{\eta} \right)^{2}
       + \tilde{\theta}_{0y}^{2} \right]} , \\
 \tilde{\theta}_{1y}
 & = & \frac{\tilde{\theta}_{0y} \left[ \left( 
       \tilde{\theta}_{0x}^{2} + \tilde{\theta}_{0y}^{2} 
       \right)^{2} + \left( \tilde{\theta}_{0x}^{2}  
       - \tilde{\theta}_{0y}^{2} \right) 
       - 2 \tilde{\theta}_{0x} 
       \left( \tilde{\theta}_{0x} - \tilde{\eta} \right) \right]}
       {\left[ \left( \tilde{\theta}_{0x}^{2} 
       + \tilde{\theta}_{0y}^{2} \right)^{2} - 1 \right]
       \left[ \left( \tilde{\theta}_{0x} - \tilde{\eta} \right)^{2}
       + \tilde{\theta}_{0y}^{2} \right]} .
\end{eqnarray}
\end{subequations}

\begin{figure}
    \epsfxsize= 9 cm
    \centerline{\epsfbox{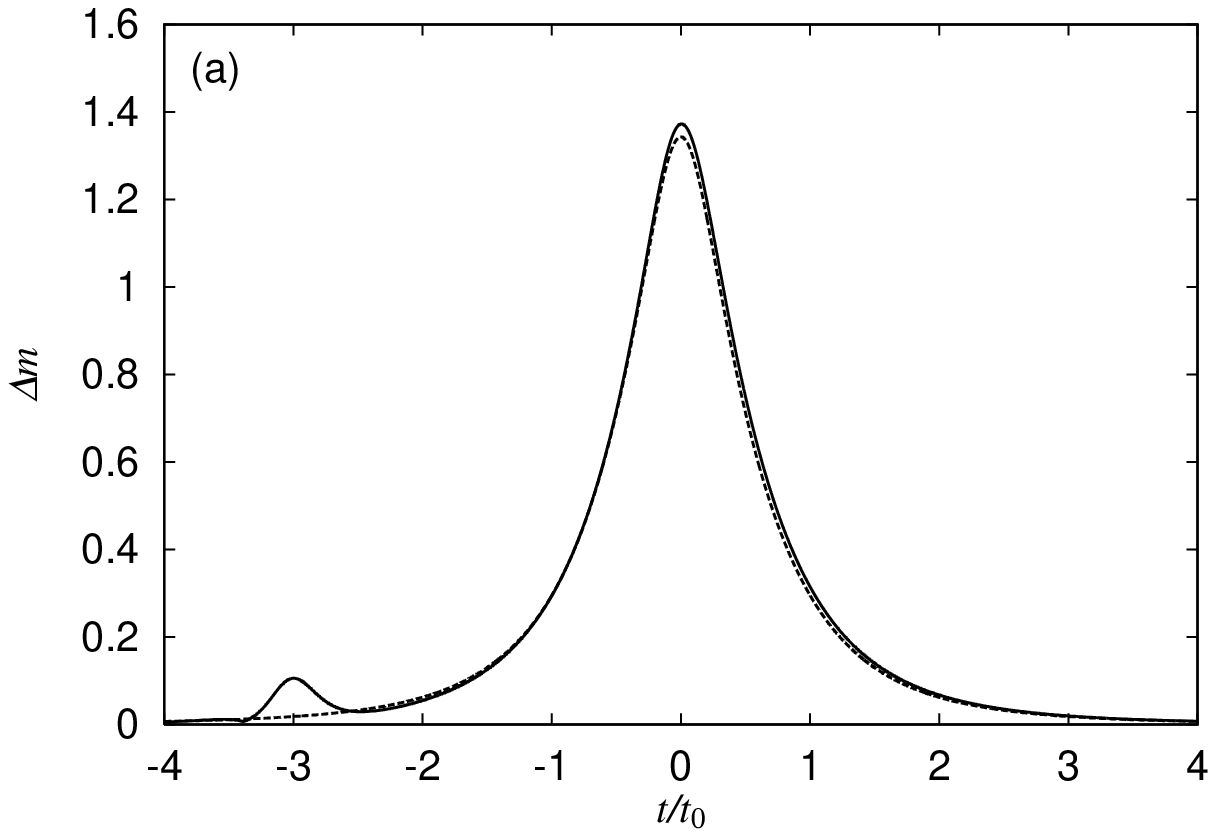}}
    \epsfxsize= 9 cm
    \centerline{\epsfbox{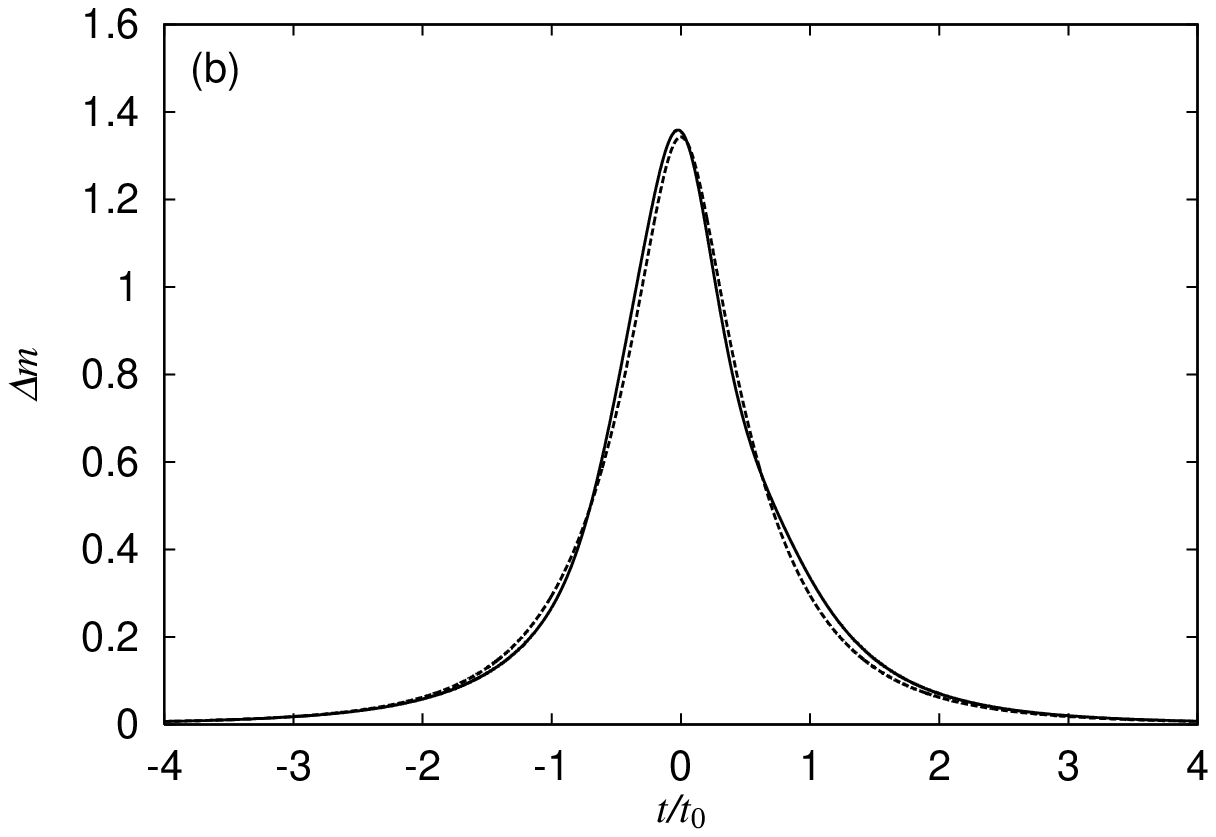}}
    \epsfxsize= 9 cm
    \centerline{\epsfbox{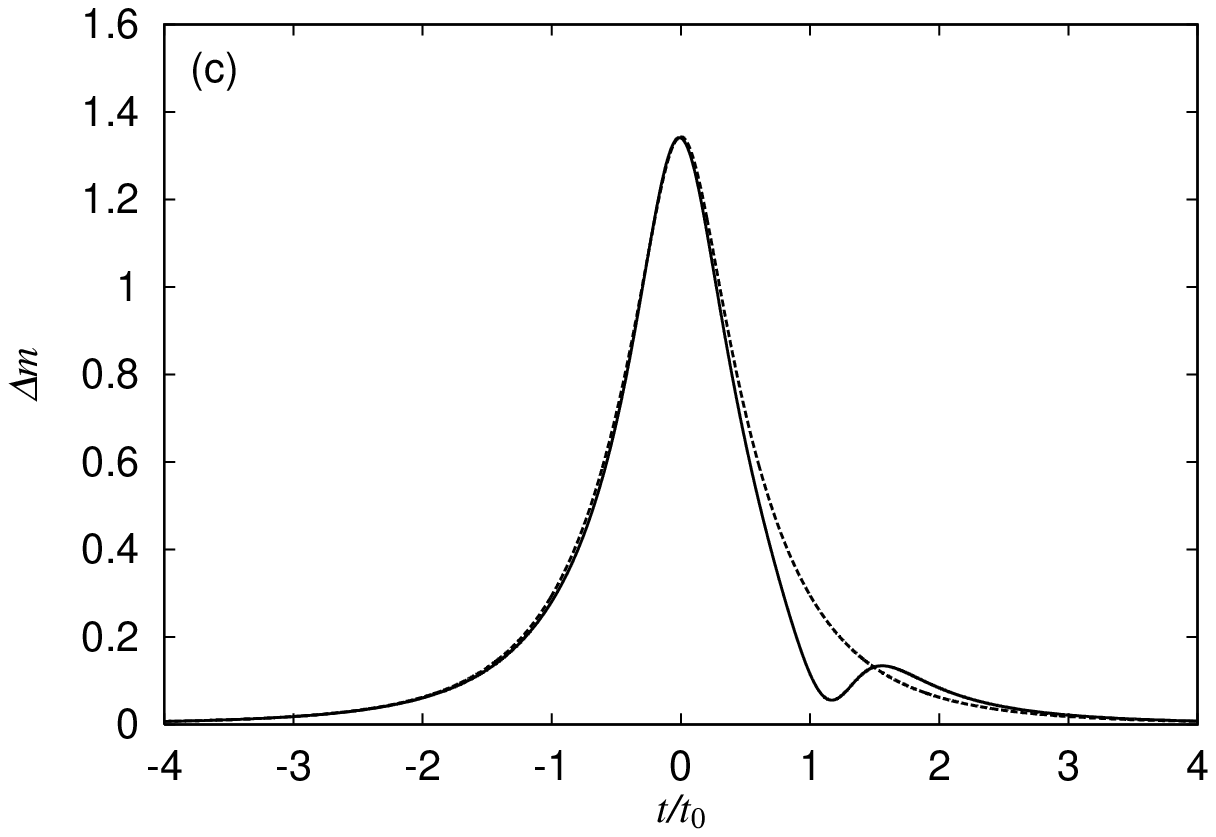}}
    \caption{Microlensing-induced light curves with (solid) and
      without (dashed) correction for the deviation from a 
      single-point-mass lens. The light curves with the correction
      are plotted for the case of the small parameter value $\mu = 0.05$
      and for angular separations of (a) $\tilde{\eta} = 0.3$,
      (b) $\tilde{\eta} = 1.0$, and (c) $\tilde{\eta} = 1.7$.
      The relative motion of the source to the lens is assumed
      to be described by $\varphi = 0$ and $p=0.3$.}
    \label{graph2}
\end{figure}

Furthermore, we can calculate the magnification by using 
Eq.~(\ref{mag}) and the inverse of the Jacobian
$\det \left( \frac{\partial \tilde{\beta}_{i}}{\partial 
\tilde{\theta}_{j}} \right)$, 
which is up to first order in $\mu$,
given by
\begin{eqnarray}
 \left[ \det \left( \frac{\partial \tilde{\beta}_{i}}
   {\partial \tilde{\theta}_{j}}\right)\right]^{-1}
 & = & \left[ 1 - \frac{1}{\left( \tilde{\theta}_{0x}^{2}
       + \tilde{\theta}_{0y}^{2} \right)^{2}}
       + \mu \frac{4 \left( \tilde{\theta}_{0x} \tilde{\theta}_{1x}
       + \tilde{\theta}_{0y} \tilde{\theta}_{1y} \right)}
       {\left( \tilde{\theta}_{0x}^{2} 
       + \tilde{\theta}_{0y}^{2} \right)^{3}} \right. 
       \nonumber \\
 & & \quad \left. 
       - \: \mu \frac{2 \left( \tilde{\theta}_{0x}^{2} 
       - \tilde{\theta}_{0y}^{2} \right) \left[
       \left( \tilde{\theta}_{0x} - \tilde{\eta} \right)^{2}
       - \tilde{\theta}_{0y}^{2} \right] + 8 \tilde{\theta}_{0x}
       \left( \tilde{\theta}_{0x} - \tilde{\eta} \right)
       \tilde{\theta}_{0y}^{2}}
       {\left( \tilde{\theta}_{0x}^{2} 
       + \tilde{\theta}_{0y}^{2} \right)^{2}
       \left[ \left( \tilde{\theta}_{0x} - \tilde{\eta} \right)^{2}
       + \tilde{\theta}_{0y}^{2} \right]^{2}} \right]^{-1}
       \nonumber \\
 & = & \left( 1 - \frac{1}{\left( \tilde{\theta}_{0x}^{2}
       + \tilde{\theta}_{0y}^{2} \right)^{2}} \right)^{-1}
       \nonumber \\
 & & \times \:
       \left[ 1 - \mu \frac{4 \left( \tilde{\theta}_{0x} 
       \tilde{\theta}_{1x} + \tilde{\theta}_{0y} 
       \tilde{\theta}_{1y} \right)}
       {\left( \tilde{\theta}_{0x}^{2} 
       + \tilde{\theta}_{0y}^{2} \right) 
       \left[ \left( \tilde{\theta}_{0x}^{2}
       + \tilde{\theta}_{0y}^{2} \right)^{2} - 1 \right]}
       \right. \nonumber \\
 & & \qquad \left.
       + \: \mu \frac{2 \left( \tilde{\theta}_{0x}^{2} 
       - \tilde{\theta}_{0y}^{2} \right) \left[
       \left( \tilde{\theta}_{0x} - \tilde{\eta} \right)^{2}
       - \tilde{\theta}_{0y}^{2} \right] + 8 \tilde{\theta}_{0x}
       \left( \tilde{\theta}_{0x} - \tilde{\eta} \right)
       \tilde{\theta}_{0y}^{2}}
       {\left[ \left( \tilde{\theta}_{0x}^{2}
       + \tilde{\theta}_{0y}^{2} \right)^{2} - 1 \right]
       \left[ \left( \tilde{\theta}_{0x} - \tilde{\eta} \right)^{2}
       + \tilde{\theta}_{0y}^{2} \right]^{2}} \right] .
       \nonumber \\
\end{eqnarray}
Therefore, we can also obtain the microlensing-induced light curves
\begin{equation}
 \Delta m = \Delta m (t)
   = 2.5 \log_{10} M \left( \beta_{x} (t), \beta_{y} (t) ; \mu \right) ,
\end{equation}
where the relative source trajectory is described by 
Eq.~(\ref{s-traj}) using the angle $\varphi$ and the impact
parameter $p$.
Some examples of the light curves with different 
values of $\tilde{\eta}$ are shown in Fig.~\ref{graph2} for $\mu = 0.05$,
$\varphi=0$ and $p=0.3$.
%
As seen in these figures, some light curves
tend to have double peaks with certain geometrical configurations.
The second peak corresponds to the effect of a planet of
mass $0.05 M_2. $ The ratio of the
height at the peaks is almost determined by the 
mass ratio $ \mu .$
Although we have restricted ourselves to the case $\mu \ll 1$
in order to exclude peculiar light curves, light 
curves with double peaks are still obtained with certain
configurations.
This is because the correction term 
$\left[ \left( \theta_{0x} - \tilde{\eta} \right)
+ \theta_{0y}^{2} \right]^{-1}$ becomes effective when
$\theta_{0x}$ is equal to $\tilde{\eta}$.
(More detailed discussions of planetary-binary lensing 
with dramatic features are given in Refs.~\citen{mp}--\citen{gg}.)

\begin{figure}
    \epsfxsize= 9 cm
    \centerline{\epsfbox{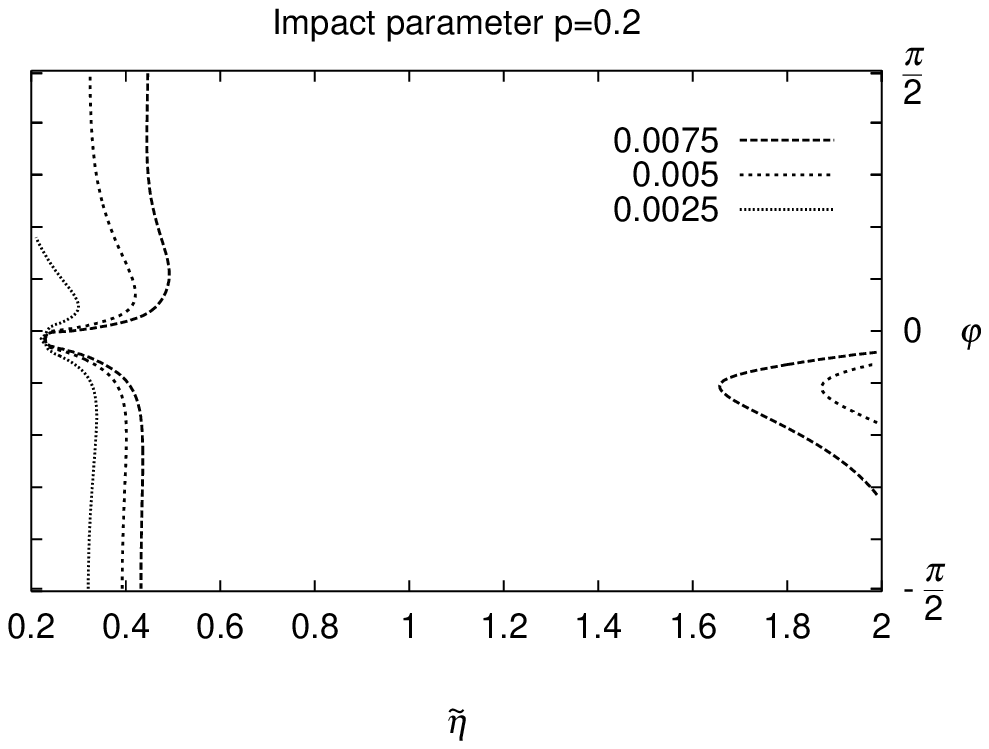}}
    \epsfxsize= 9 cm
    \centerline{\epsfbox{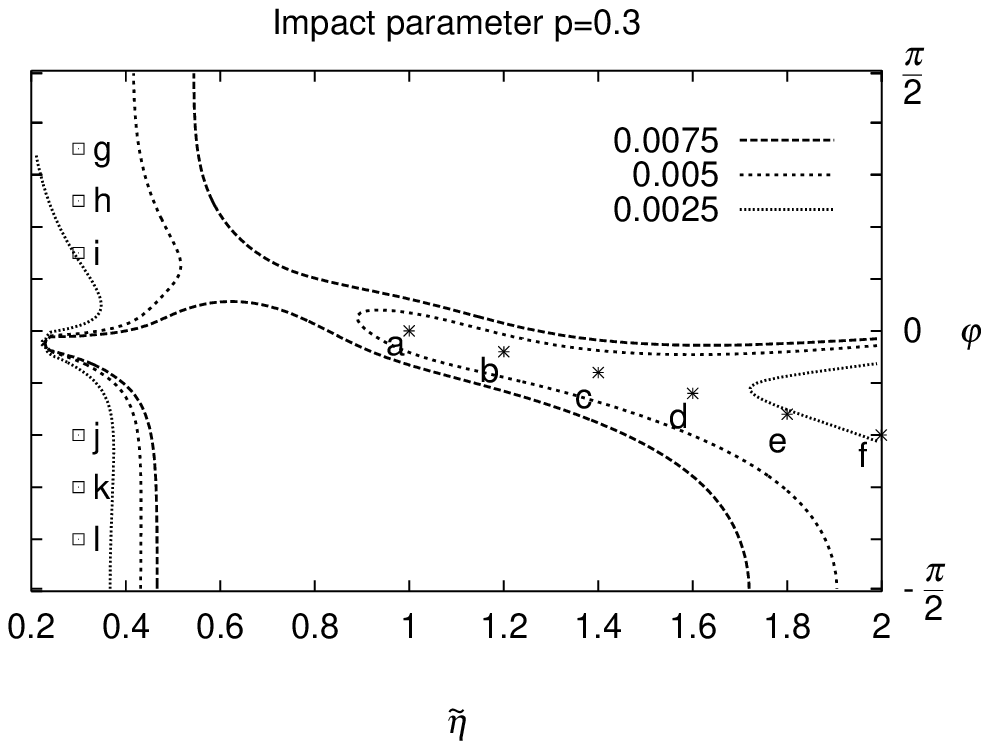}}
    \epsfxsize= 9 cm
    \centerline{\epsfbox{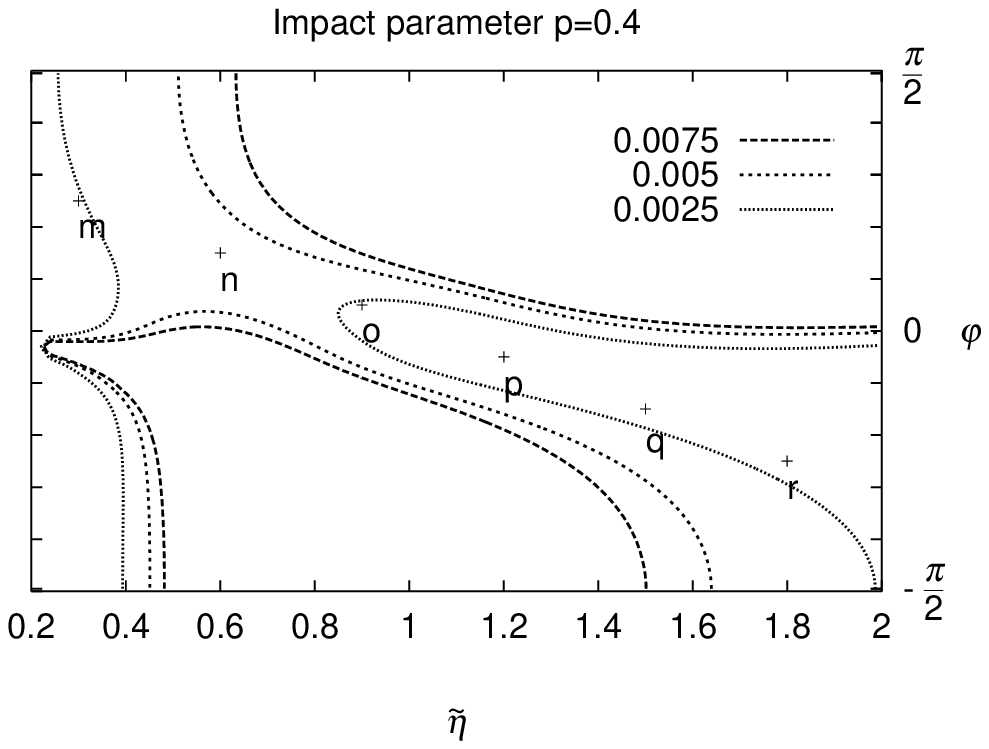}}
    \caption{Contours of $\delta$ in
             $\tilde{\eta}$-$\varphi$ space for the impact
             parameters $p=0.2$, $0.3$ and $0.4$. The small
             parameter $\mu$ is assumed to be 0.05.
             The attached labels `{a}'--`{r}' indicate
             the parameters used in Fig.~10 (see text).}
    \label{contour}
\end{figure}

Furthermore, it is interesting to investigate the
configurations in which the Lorentzian curves, i.e.,
almost time-symmetric light curves arise.
For this purpose, we consider the integral of the square of
the magnitude difference from the corresponding
single-point-mass lens: 
$\delta \equiv \int_{- \infty}^{+ \infty} ( \Delta m - \Delta m_{0} )^2 dt$.
The quantity $\delta$ indicates the criterion of the 
deviation from the light curve due to the single-point-mass lens.
Figure \ref{contour} displays the contours of $\delta$ in 
the $\tilde{\eta}$-$\varphi$ space
for different impact parameters.
Of course, as the quantity $\delta$ becomes smaller, 
the light curve moves closer to that due to the 
single-pint-mass lens. In fact, we can find 
almost time-symmetric light curves
in the domains where $\delta$
is smaller than $\sim 0.005$, as seen in Fig.~\ref{graph2}(b).
Such domains tend to become larger as the impact parameter $p$
increases. The same tendency can also be derived by making the 
mass ratio $\mu$ small.

\section{Rotating objects}
\label{kerr}

Several relativistic effects also causes corrections
to the point-mass lens. 
We  only consider the  dragging effect of inertial frames
arising from a rotating object, 
since other spherical post-Newtonian terms never
induce asymmetry in light curves.
This additional effect is described with the 
spin angular momentum ${\mbf J}.$ \ 
\begin{figure}[t]
    \epsfxsize= 8 cm
    \centerline{\epsfbox{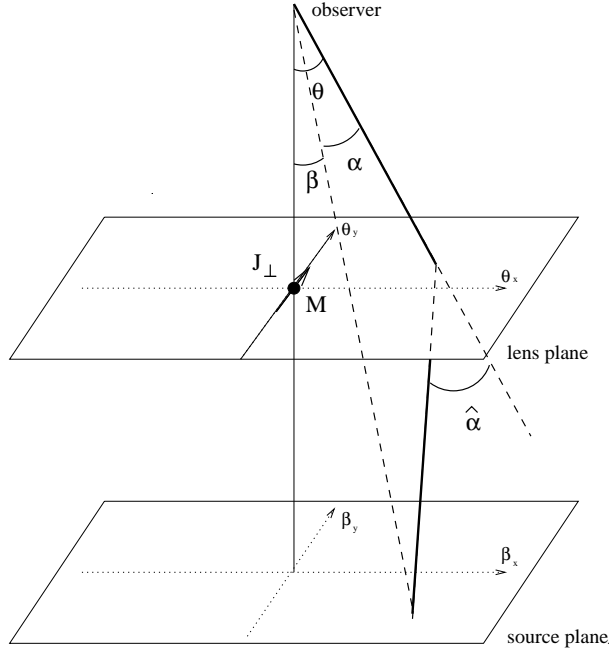}}
    \caption{Geometry of lensing by the rotating object considered in \S 3.}
    \label{lens3}
\end{figure}
We consider the projection
of the angular momentum of the rotating object on to the lens
plane and define the $(\theta_{x},\theta_{y})$ coordinate
system so that the $\theta_{y}$-axis is oriented parallel to 
the projected angular momentum ${\mbf J}_{\perp}$
(see Fig.~\ref{lens3}).
The deflection angles
$(\hat{\alpha}_{x},\hat{\alpha}_{y})$ are written as~\cite{sari} 
\begin{subequations}
\begin{eqnarray}
 \hat{\alpha}_{x} 
   & = & \frac{4GM}{c^2 D_{L}} \frac{\theta_{x}}
         {\theta_{x}^{2} + \theta_{y}^{2}}
         + \frac{4GJ_{\perp}}{c^3 D_{L}^{2}}
         \frac{\theta_{x}^{2} - \theta_{y}^{2}}
         {\left( \theta_{x}^{2} + \theta_{y}^{2} 
         \right)^{2}} , \\
 \hat{\alpha}_{y} 
   & = & \frac{4GM}{c^2 D_{L}} \frac{\theta_{y}}
         {\theta_{x}^{2} + \theta_{y}^{2}}
         + \frac{4GJ_{\perp}}{c^3 D_{L}^{2}}
         \frac{2 \theta_{x} \theta_{y}}
         {\left( \theta_{x}^{2} + \theta_{y}^{2} 
         \right)^{2}} .
\end{eqnarray}
\end{subequations}
Hence, we have
\begin{subequations}
\begin{eqnarray}
 \alpha_{x} 
   & = & \frac{4GM}{c^2} \frac{D_{LS}}{D_{L} D_{S}} \frac{\theta_{x}}
         {\theta_{x}^{2} + \theta_{y}^{2}}
         + \frac{4GJ_{\perp}}{c^3} \frac{D_{LS}}{D_{L}^{2} D_{S}}
         \frac{\theta_{x}^{2} - \theta_{y}^{2}}
         {\left( \theta_{x}^{2} + \theta_{y}^{2} 
         \right)^{2}} , \\
 \alpha_{y} 
   & = & \frac{4GM}{c^2} \frac{D_{LS}}{D_{L} D_{S}} \frac{\theta_{y}}
         {\theta_{x}^{2} + \theta_{y}^{2}}
         + \frac{4GJ_{\perp}}{c^3} \frac{D_{LS}}{D_{L}^{2} D_{S}}
         \frac{2 \theta_{x} \theta_{y}}
         {\left( \theta_{x}^{2} + \theta_{y}^{2} 
         \right)^{2}} . 
\end{eqnarray}
\end{subequations}
Therefore, using the quantities normalized by the Einstein
radius, the lens equation becomes
\begin{subequations}
\begin{eqnarray}
 \tilde{\beta}_{x}
 & = & \tilde{\theta}_{x} - \frac{\tilde{\theta}_{x}}
       {\tilde{\theta}_{x}^{2} + \tilde{\theta}_{y}^{2}}
       - \gamma \frac{\tilde{\theta}_{x}^{2} - \tilde{\theta}_{y}^{2}}
       {\left( \tilde{\theta}_{x}^{2} + \tilde{\theta}_{y}^{2}
       \right)^{2}} , \\
 \tilde{\beta}_{y}
 & = & \tilde{\theta}_{y} - \frac{\tilde{\theta}_{y}}
       {\tilde{\theta}_{x}^{2} + \tilde{\theta}_{y}^{2}}
       - \gamma \frac{2 \tilde{\theta}_{x} \tilde{\theta}_{y}}
       {\left( \tilde{\theta}_{x}^{2} + \tilde{\theta}_{y}^{2}
       \right)^{2}} ,
\end{eqnarray}
\end{subequations}
where $\gamma$ is given by 
\begin{equation}
 \gamma = \frac{1}{\theta_{E}} \cdot \frac{J_{\perp}}{Mc D_{L}} .
\end{equation}
It is interesting that this lens equation is the same as that of the 
two-point-mass lenses for the close binary system, $l \ll r_{E}$, 
with the correspondence 
$\gamma \leftrightarrow \varepsilon$.
Hence, the same asymmetric light curves are obtained.
Furthermore, it is impossible to distinguish
two such corrections.
However, the parameter $\gamma$ is quite small.
For example,
we consider a Kerr black hole. The angular momentum
is $ J \sim G M^2/c $, so that
we have $ \gamma \sim ( G M D_S/(c^2 D_{L}D_{LS}))^{1/2} \ll  1 .$

\section{Quantitative estimate}
\label{n-estimates}

\begin{figure}[t]
    \epsfxsize= 8 cm
    \centerline{\epsfbox{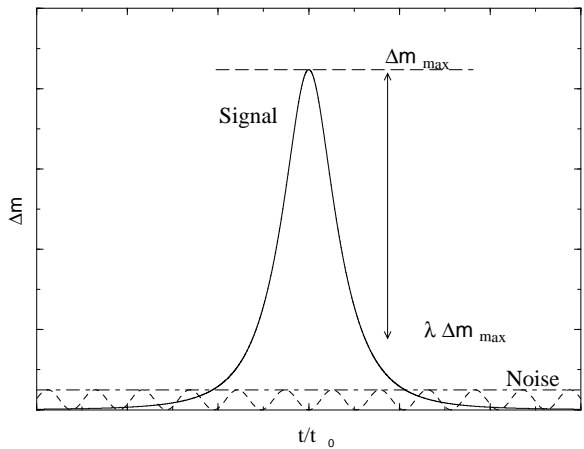}}
    \caption{Illustration of the discussion given in \S 4.
             A schematic light curve with noise.}
    \label{fig:s-n}
\end{figure}

In order to estimate the asymmetry in the light curves quantitatively,
we now introduce the notion of `skewness' from statistics.\cite{stat} \ 
In statistics, the skewness for any distribution function $f(t)$
is defined as
\begin{equation}
\label{skew-d}
  \mbox{skewness} = \frac{1}{N} \int_{- \infty}^{\infty} \!
                \left( \frac{t - \mu}{\sigma} \right)^{3}
                f(t) dt , 
\end{equation}
where $N$ is the normalization factor, $\mu$ the mean,
and $\sigma$ the standard deviation. However, there
is a problem in utilizing the skewness exactly in this form.
To see this, we consider a single-point-mass lens. The light curve 
is then given by
\begin{equation}
  \Delta m = 2.5 \log_{10} \left[ \left|
    \frac{\left( \tilde{\theta}_{x_{+}}^{2}
    + \tilde{\theta}_{y_{+}}^{2} \right)^2}
    {\left( \tilde{\theta}_{x_{+}}^{2} + \tilde{\theta}_{y_{+}}^{2}
    \right)^{2} - 1} \right| + \left|
    \frac{\left( \tilde{\theta}_{x_{-}}^{2}
    + \tilde{\theta}_{y_{-}}^{2} \right)^2}
    {\left( \tilde{\theta}_{x_{-}}^{2} + \tilde{\theta}_{y_{-}}^{2}
    \right)^{2} - 1} \right| \right] .
\end{equation}
If we set $\tilde{\beta}_{x} = t/t_{0}$ and 
$\tilde{\beta}_{y} = \mbox{const.}$ and consider the case that
the time $t$ approaches infinity, then we have
\begin{subeqnarray}
 \tilde{\theta}_{x_{+}} & \rightarrow & 
   \tilde{\beta}_{x} = t/t_{0} \; \left( \rightarrow \infty \right) , \\
 \tilde{\theta}_{y_{+}} & \rightarrow & 
   \tilde{\beta}_{y} = \mbox{const.} , \\
 \tilde{\theta}_{x_{-}} & \rightarrow & 0 , \\
 \tilde{\theta}_{y_{-}} & \rightarrow & 0 .
\end{subeqnarray}
It follows that at large $t$,
\begin{eqnarray}
 \Delta m & \simeq & 2.5 \log_{10} \left[  \left(
     1 - \frac{1}{t^{4}} \right)^{-1} \right] \nonumber \\
 & \simeq & \frac{2.5}{\ln 10} \cdot \frac{1}{t^{4}} .
\end{eqnarray}
From this, we find that the integral
\begin{equation}
  \int_{\mu}^{\infty} \! t^{n} \Delta m(t) dt
\end{equation}
diverges if $n \geq 3$. Therefore, we cannot use 
Eq.~(\ref{skew-d}) itself. Nevertheless, since
actually observed light curves necessarily include noise,
it seems that the integral to infinity is meaningless.
With this consideration, we define the skewness
for the restricted region of a light curve 
$\Delta m   > \lambda \Delta m_{\mbox{max}}$ , where
$\Delta m_{\mbox{max}}$ is the maximum value.
Here we have introduced the cutoff $\lambda$ $(0 < \lambda < 1),  $
and the usual skewness corresponds to $ \lambda =0. $
The cutoff will naturally appear in the actual data,
e.g., the region $ \Delta m < \lambda \Delta m_{\mbox{max}}$
may be meaningless due to  the noise level (see Fig.~\ref{fig:s-n}).
In microlensing-induced light curves, the
bottom level is constant, but the maximum value may have
some uncertainty as 
$ ( 1  \pm \lambda )\Delta m_{\mbox{max}} .$

\begin{figure}[t]
    \epsfxsize= 8 cm
    \centerline{\epsfbox{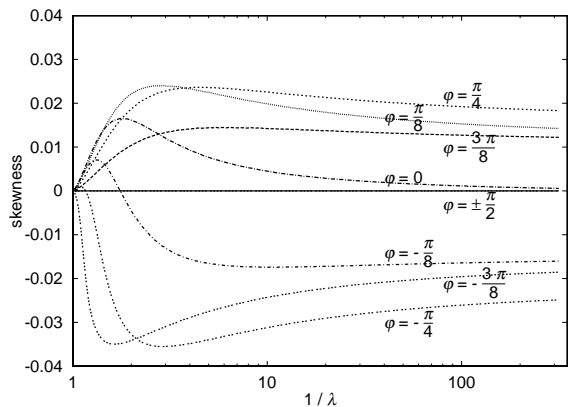}}
    \caption{Dependence of the skewness on the angle
             $\varphi$ in the close binary case. 
             The trajectories have the same impact parameter,
             $p=0.3$, and the light curves are calculated 
             for the small parameter value 
             $\varepsilon=0.1$.}
    \label{fig:skp-a}
\end{figure}

\begin{figure}[t]
    \epsfxsize= 8 cm
    \centerline{\epsfbox{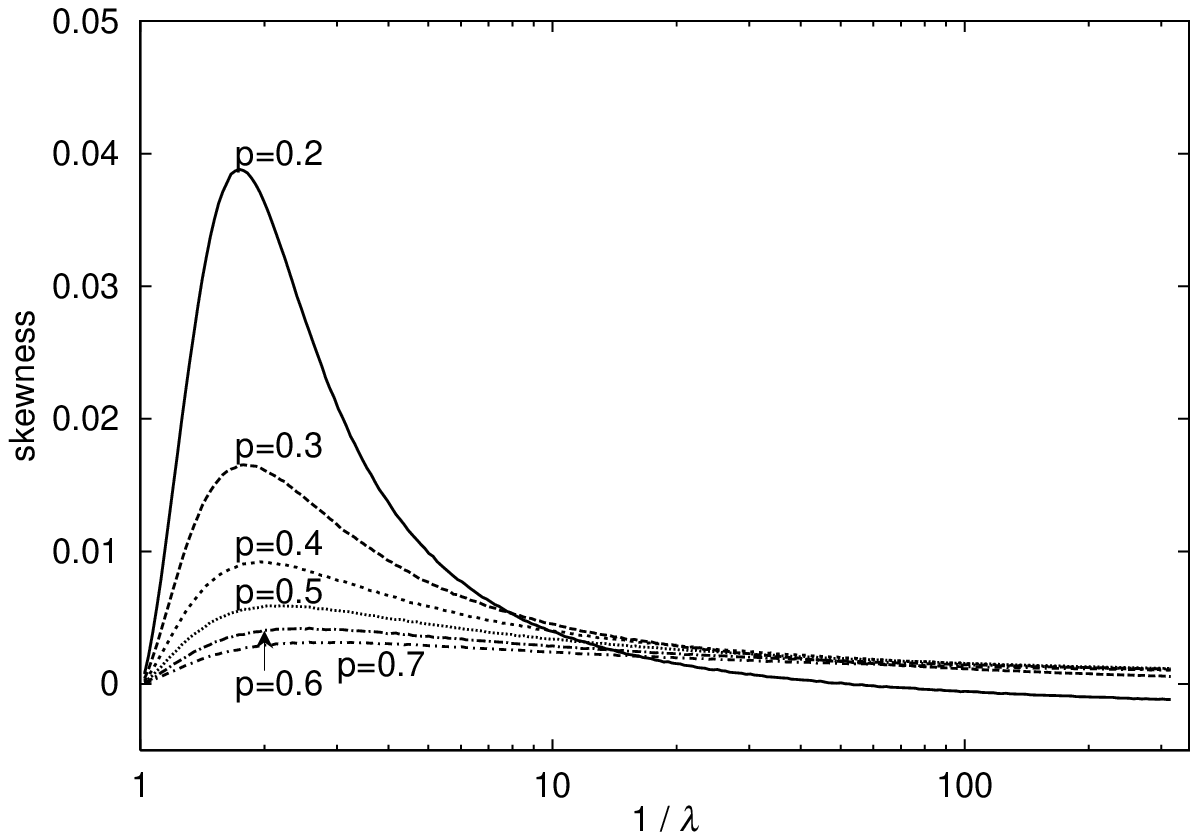}}
    \caption{Dependence of the skewness on the impact parameter
             $p$ in the close binary case. The trajectories 
             have the same angle, $\varphi=0$, and the light 
             curves are calculated for the small parameter value 
             $\varepsilon=0.1$.}
    \label{fig:ska-p}
\end{figure}

\begin{figure}[t]
    \epsfxsize= 8 cm
    \centerline{\epsfbox{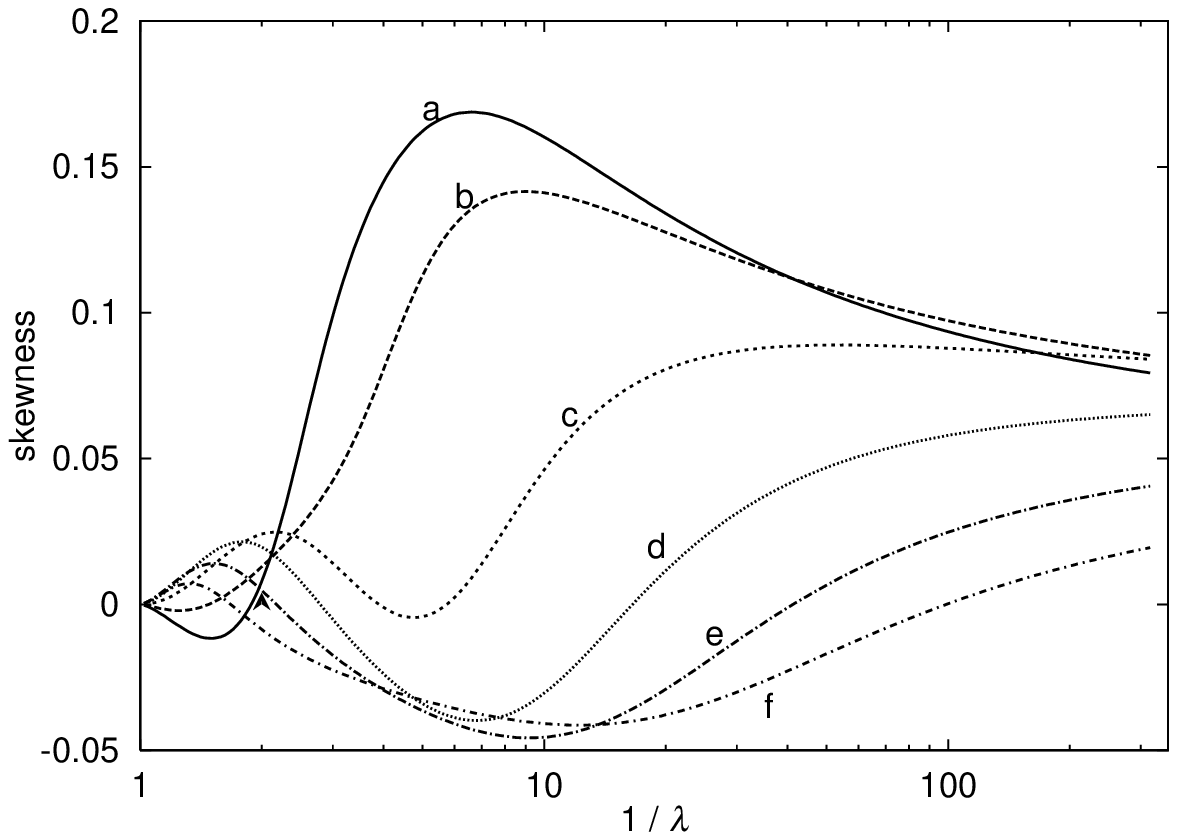}}
    \epsfxsize= 8 cm
    \centerline{\epsfbox{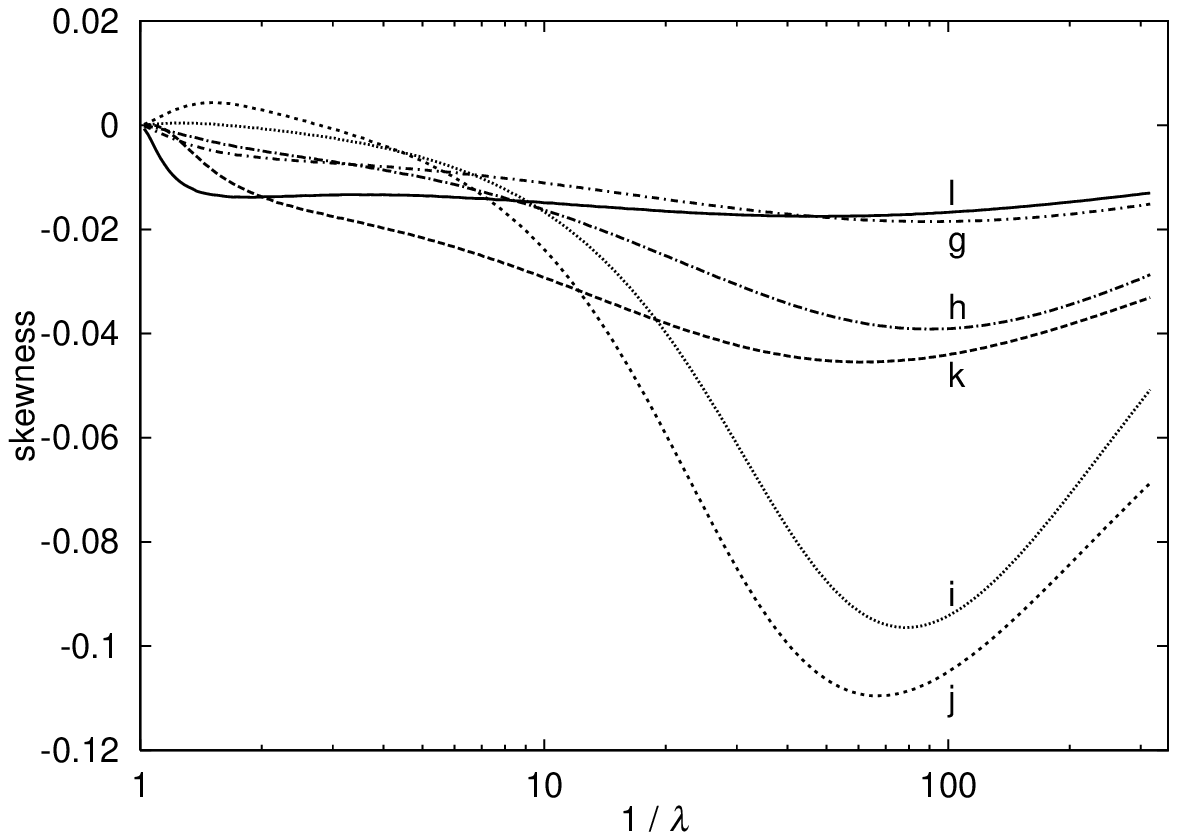}}
    \epsfxsize= 8 cm
    \centerline{\epsfbox{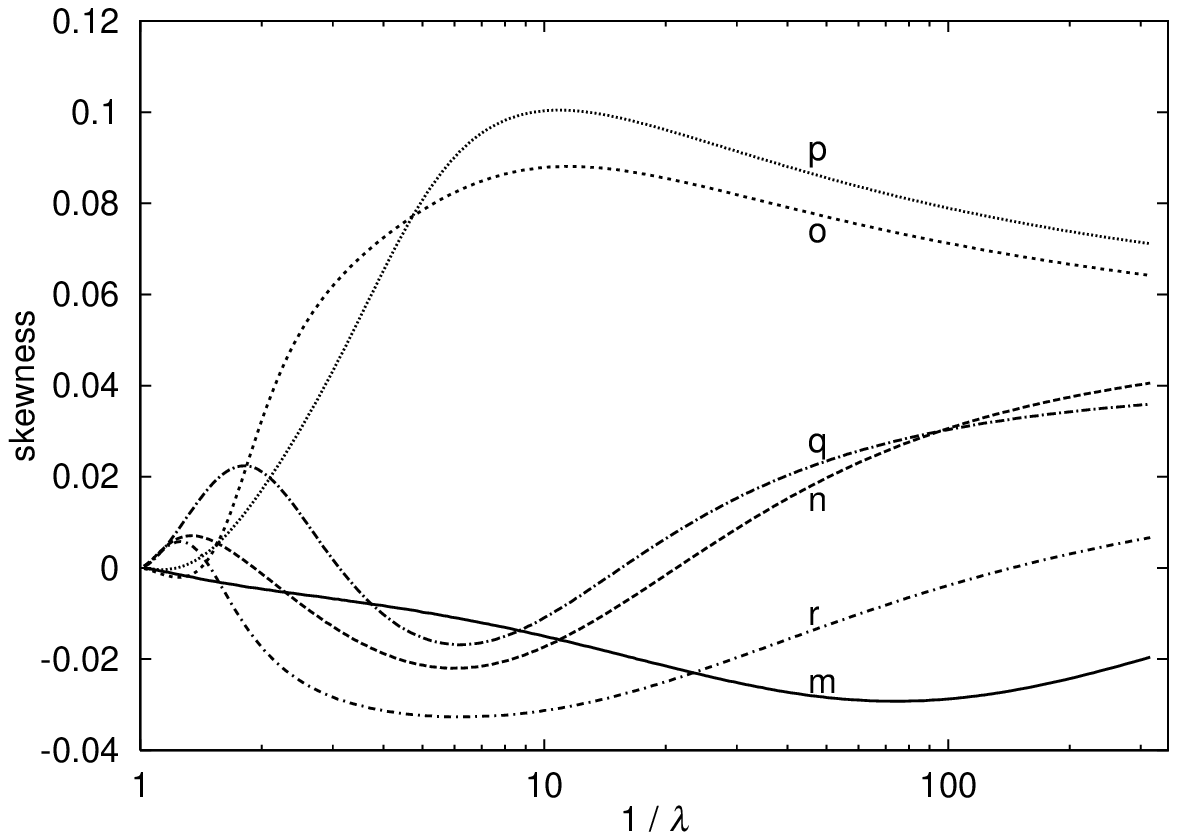}}
    \caption{The skewness of light curves in the planetary
             system. The curves labeled `{a}'--`{r}'
             correspond to the points labeled `{a}'--`{r}', 
             respectively, in Fig.~5. The light curves are calculated
             for the small parameter value $\mu = 0.05$.}
    \label{fig:bs}
\end{figure}

\begin{figure}[t]
    \epsfxsize= 8 cm
    \centerline{\epsfbox{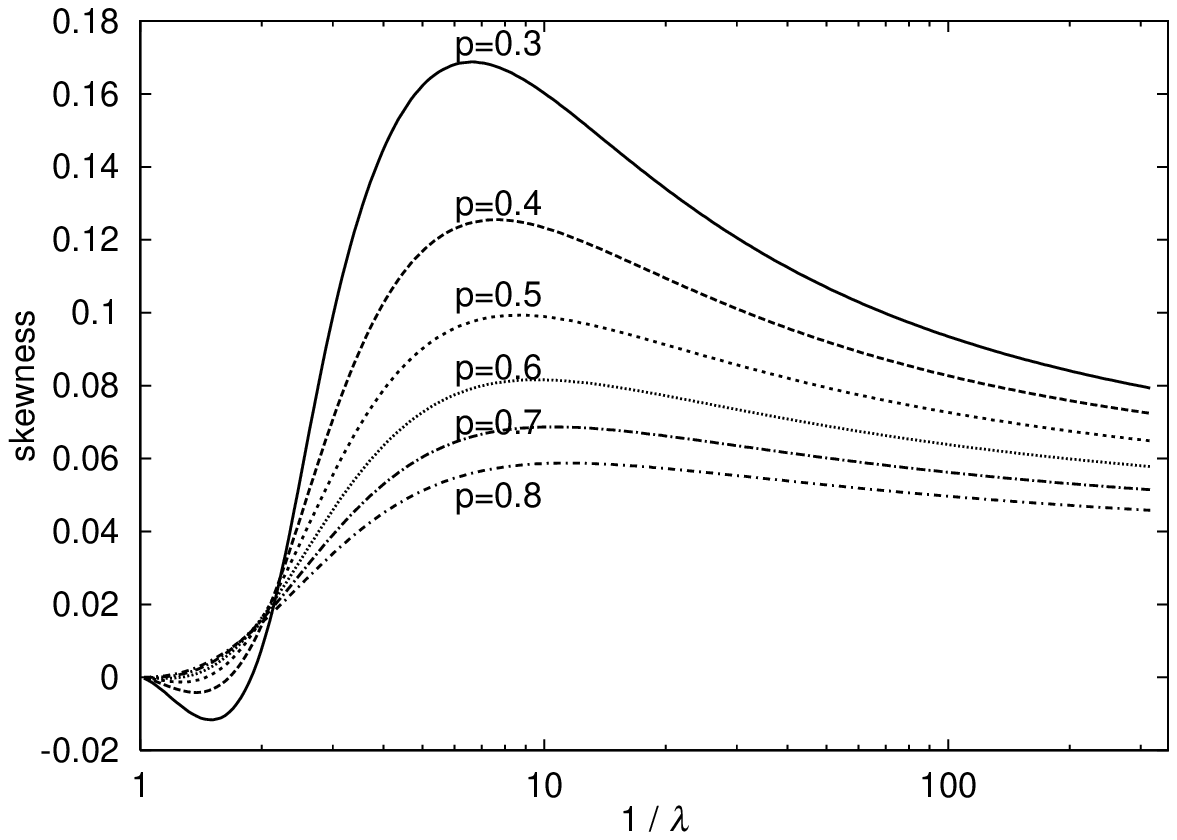}}
    \caption{Dependence of the skewness on the impact parameter
             $p$ in the planetary case. The light curves correspond 
             to the point `{a}' in Fig.~5, 
             and are calculated for the small parameter value 
             $\mu=0.05$.}
    \label{fig:skba-p}
\end{figure}

We apply this tool to the case of almost identical light curves,
which are caused from quite different physical situations,
that is, a close binary and a planetary system.
The skewness of these light curves are respectively displayed as 
functions of $1/\lambda$
in Figs.~\ref{fig:skp-a}--\ref{fig:skba-p}.
Figures \ref{fig:skp-a} and \ref{fig:ska-p} display the 
dependence of the skewness of the close binary light curves
on the angle $\varphi$ and the impact 
parameter $p$, respectively.
The light curve given in Fig.~\ref{graph1} corresponds to 
the curves in Figs.~\ref{fig:skp-a} and \ref{fig:ska-p}.
On the other hand, Figs.~\ref{fig:bs}
and \ref{fig:skba-p} display the skewness of the light curves in
the planetary system. In the case of the planetary system,
the almost time-symmetric light curves are derived only
under certain configurations, as indicated in \S \ref{ps}.
Figure \ref{fig:bs} displays the skewness of the light curves under
such configurations. The curves labeled by `{a}'--`{r}'
correspond to the points labeled by `{a}'--`{r}', respectively,
in Fig.~\ref{contour}.
Figure \ref{fig:skba-p} displays the impact parameter
dependence of the curves corresponding to the point `{a}'.
The light curve given in Fig.~\ref{graph2}(b) corresponds
to the curve `{a}' in Fig.~\ref{fig:bs} and a curve in 
Fig.~\ref{fig:skba-p}.

As is expected, we have zero skewness when 
$\varphi=\pm \frac{\pi}{2}$ in Fig.~\ref{fig:skp-a}.
However, when $\varphi$ is different from $\pm \frac{\pi}{2}$,
we have comparable values of the skewness.
This demonstrates the useful aspect of the method using the 
skewness. Nevertheless, the skewness becomes smaller
as the impact parameter increases, as seen in 
Figs.~\ref{fig:ska-p} and \ref{fig:skba-p}.
Thus, the usefulness of the method using the skewness
depends mainly on the impact parameter.
The skewness with respect to different values
of $\lambda$ fully shows the asymmetric features of
the light curves.
The absolute values of the skewness depend on the
the small parameters $\varepsilon $ and $ \mu $
(i.e., the binary separation and the mass ratio).
Furthermore, it should be noted that the absolute values 
of the skewness have a maximum at 
$1/\lambda \sim 2$--$4$ and decrease monotonously in the 
close binary case, while in the planetary case the skewness
indicates different behavior, peaks at larger
values of $1/\lambda$, and so on.
Therefore,  we may discriminate the underlying cases for the
asymmetry by the $\lambda$-dependent skewness 
for a good signal-to-noise ratio beyond $ \sim 10 $.

\section{Summary and discussion}
\label{s-d}

We have studied distortion in microlensing-induced light curves
which are considered to be curves of single-point-mass lenses at a 
first glance. In particular, we have attributed factors inducing
the distortion to lenses themselves and considered two sorts of 
corrections: corrections due to deviation from the Newtonian
gravitational potential $\phi = - GM/r$ and corrections due to
general relativistic effects of dragging of inertial frames
arising from a rotating object. For simplicity, we have 
discussed two extreme cases in two-point-mass lenses
for the corrections of the potential; one is the close binary
case in which $l \ll r_{E}$, and the other is the planetary system
case in which
$M_{1} \ll M_{2}$. Moreover, we have considered 
corrections up to the post-Newtonian order for the effect 
of dragging of inertial frames. From this, we found
the same time-asymmetric light curves as in the 
two-point-mass lenses where $l \ll r_{E}$.
Furthermore, we have introduced the cutoff dependent skewness
and estimated the asymmetry in the light curves quantitatively.
In particular, we showed the clear difference in the 
skewness for almost similar light curves.

Here we make a comment on the additional factor
for the asymmetry in the binary system.
In this paper, we have assumed that the lens systems 
of the two point masses are fixed, but 
each star constituting the binary
revolves around the center of mass. Therefore,
the rotational effect may also appear.
However, if the rotation period 
$T \sim (l^3/(G(M_1 +M_2) )^{1/2}$ 
of the binary is much larger than the typical time
scale $t_{0}$ of a microlensing event, our consideration
of the fixed lens systems would be appropriate.  
For the extreme case $T \ll t_{0}$, the time-averaged 
gravitational potential
which is projected on to the lens plane may be regarded
approximately as that of a single-point-mass lens or
a fixed two-point-mass lens with $l \ll r_{E}$ if the
lens is compact.  Therefore,
more complicated variations of light curves, corresponding
to the phase, are  expected only  if $T \sim t_{0}$.

It is very interesting whether the time-asymmetric features
of light curves discussed in this paper will actually be detected
by the projects
(MOA, etc.) that are now in progress.

\section*{Acknowledgements}
This work was supported in part
by a Grant-in-Aid for Scientific Research Fund of
the Ministry of Education, Science, Sports and Culture of Japan
(08640378).



\begin{thebibliography}{99}
\bibitem{microl}
  See, e.g., B.~Paczy\'nsky,
  Ann.~Rev.~Astr.~Ap. {\bf 34} (1996), 419;
  R.~Narayan and M.~Bartelmann,
  {\it Lectures on Gravitational Lensing}, 
  preprint: astro-ph/9606001.
\bibitem{cr}
  K.~Chang and S.~Refsdal,
  Nature {\bf 282} (1979), 561.
\bibitem{gott}
  J.~R.~Gott,
  Astrophys. J. {\bf 243} (1981), 140.
\bibitem{gould}
  A.~Gould,
  Astrophys. J. {\bf 392} (1992), 442.
\bibitem{alcock}
  C.~Alcock, R.~A.~Allsman, D.~Alves, et al.,
  Astrophys. J. {\bf 454} (1995), L125.
\bibitem{gg}
  B.~S.~Gaudi and A.~Gould,
  Astrophys. J. {\bf 482} (1997), 83.
\bibitem{mp}
  S.~Mao and B.~Paczy\'nski,
  Astrophys. J. {\bf 374} (1991), L37.
\bibitem{gl}
  A.~Gould and A.~Loeb,
  Astrophys. J. {\bf 396} (1992), 104.
\bibitem{bf}
  A.~D.~Bolatto and E.~E.~Falco,
  Astrophys. J. {\bf 436} (1994), 112.
\bibitem{br}
  D.~P.~Bennett and S.~H.~Rhie,
  Astrophys. J. {\bf 472} (1996), 660.
\bibitem{gg}
  B.~S.~Gaudi and A.~Gould,
  Astrophys. J. {\bf 486} (1997), 85.
\bibitem{sw}
  P.~Schneider and A.~Weiss,
  Astron. Astrophys. {\bf 164} (1986), 237.
\bibitem{sari}
  S.~O.~Sari,
  Astrophys. J. {\bf 462} (1996), 110.
\bibitem{stat}
  See, e.g., W.~H.~Press et al.,
  {\it Numerical Recipes in Fortran}
  (Cambridge University Press, 1992).
\end{thebibliography}
\end{document}